# Experimental Validation of a Dynamic Equivalent Model for Microgrids

F. Conte, *Senior Member, IEEE*, F. D'Agostino, *Member, IEEE*, S. Massucco, *Senior Member, IEEE*, F. Silvestro, *Senior Member, IEEE*, C. Bossi, M. Cabiati

***Abstract*—The goal of this paper is the experimental validation of a gray-box equivalent modeling approach applied to microgrids. The main objective of the equivalent modeling is to represent the dynamic response of a microgrid with a simplified model. The main contribution of this work is the experimental validation of a two-step process, composed by the definition of a nonlinear equivalent model with operational constraints, adapted to the microgrid environment, and the identification procedure used to define the model parameters. Once the parameters are identified, the simplified model is ready to reproduce the microgrid behavior to voltage and frequency variations, in terms of active and reactive power exchanges at the point of common coupling. To validate the proposed approach, a set of experimental tests have been carried out on a real LV microgrid considering different configurations, including both grid-connected and islanded operating conditions. Results show the effectiveness of the proposed technique and the applicability of the model to perform dynamic simulations.**

***Index Terms*— dynamic equivalents, microgrids, gray-box modeling, system identification, experimental results**

## NOMENCLATURE

| | |
|---|---|
| $T'_{ds}$ | direct-axes time constant of the equivalent Synchronous Machine (SM) [s]. |
| $H_s$ | inertia time constant of the equivalent SM [s]. |
| $x_s$ | steady-state reactance of the equivalent SM [pu]. |
| $x'_s$ | transient reactance of the equivalent SM [pu]. |
| $t_{ms}$ | mechanical torque of the equivalent SM [pu]. |
| $e_f$ | field voltage of the equivalent SM [pu]. |
| $D$ | damping factor of the equivalent SM [pu]. |
| $S_s^n$ | nominal apparent power of the equivalent SM [VA]. |
| $\Omega^n$ | nominal angular velocity of the equivalent SM [rad/s]. |
| $P_Z$ | active power absorbed by constant impedance component of the static load [W]. |
| $Q_Z$ | reactive power absorbed by constant impedance component of the static load [var]. |
| $P_I$ | active power absorbed by constant current component of the static load [W]. |
| $Q_I$ | reactive power absorbed by constant current component of the static load [var]. |
| $P_P$ | active power absorbed by constant power component of the static load [W]. |
| $Q_P$ | reactive power absorbed by constant power component of the static load [var]. |

## I. INTRODUCTION

DURING the last decades, the huge installation of distributed generation (DG) has drastically transformed the electric power system. One of the most significant differences from the past is the transformation of distribution networks, which have become active components of the power system. Indeed, they include generation and many inverter-interfaced devices, which can potentially implement dynamic responses to external disturbances. In this context, low voltage microgrids have become of major interest as one of the key elements of the smart grid infrastructures [1]. Microgrids can be used to enable the integration of distributed energy resources (DERs). For example, in [2], control methods to prevent collapse in mixed-source microgrids are evaluated. In [3] microgrids with DERs are controlled in transient overload conditions. Microgrids can be also used to provide services and flexibility to the distribution grid [4]-[5]. Moreover, several studies illustrate the impact of microgrids in improving the overall system reliability [6]-[8]. Finally, if required, microgrids can also operate in islanded mode, whereas proper control and management strategies are provided [9]-[10].

To realize consistent transient stability studies, it is therefore required to include in the power system models, the dynamic behavior of active distribution networks (ADNs) and microgrids. This issue is highly challenging, since the distribution system is complex, extremely large, and a detailed knowledge is not always available. Therefore, to build a highly detailed model of the distribution system, to be included in the overall power system model, is not possible; moreover, it would result to be useless, since the simulation of such an extremely large model could be hard and, in any case, not compatible with real time analysis. These considerations motivate the development of several approaches in defining reduced order and equivalent models for ADNs and microgrids.

Model order reduction techniques are focused on the definition the minimum set of state variables required to

F. Conte, F. D'Agostino, S. Massucco and F. Silvestro are with the Department of Electrical, Electronic, Telecommunication Engineering and Naval Architecture of the University of Genova, Genova, Italy (e-mail: fr.conte@unige.it, fabio.dagostino@unige.it, stefano.massucco@unige.it federico.silvestro@unige).

C. Bossi and M. Cabiati are with Ricerca Sistema Energetico, Milano, Italy (e-mail: claudio.bossi@rse-web.it, mattia.cabiati@rse-web.it).

reproduce the original system with an adequate fidelity. These techniques have been applied to microgrids in [11]-[13]. Moreover, in [14]-[16] reduced order models are adopted for stability analysis of isolated microgrids. In [17] a reduced order linear model is used to control a microgrid. The idea beyond the equivalent modeling approach is define and identify an Equivalent Dynamic Model (EDM), usually composed by a set of differential equations, able to reproduce the system dynamics [18]-[20]. Generally, measurement-based methods [21]-[22] are preferred to other approaches since they do not require a detailed a priori knowledge of the real system. In [23]-[25] artificial neural networks (ANNs) are used to identify EDMs for ADNs and microgrids, which can be considered as a special case of ADNs. Prony analysis is applied in [26] and [27]. In [28] a multi-input-multi-output (MIMO) transfer function is adopted to represent a microgrid.

All mentioned methods use the so called *black-box* approach, which does not assume any pre-established form of models. Differently, in the *gray-box* approach, a model structure is selected using the available information about the target system. Gray-box models are recommended in [19] and [29] since black-box ones require more measurements to be identified, and are less adaptable to different network configurations.

A gray-box model for ADNs is proposed in [30], and validated by simulations in [31]. An improved version is introduced in [32]. A similar approach is illustrated in [33] for grid-connected microgrids. In these papers, the proposed model is nonlinear, and the parameters identification is carried out using nonlinear optimization. In [34], a similar approach is proposed, where the main advantage is that the parameter identification procedure is realized by adding operational and modeling constraints, in accordance with the basic knowledge of the ADN. This is also a key aspect to reduce the size of the solution space and to obtain the convergence of the optimization algorithm, since the problem is highly nonlinear and non-convex. Results in [34] prove the suitability of the model to be adapted to different configurations of the network by modifying a limited set of scenario parameters. Moreover, a comparison of the results obtained with the EDM and a recurrent ANN trained with the same dataset, highlights that, in the majority of cases, the EDMs clearly outperform ANNs. In [35]-[38], modeling constraints are included in the identification of an EDM for ADNs that includes inverter-interfaced generators.

The mentioned gray-box approaches have been validated by simulation analysis, using highly detailed models to represent the real field. To the author knowledge, no experimental validation has been provided for the existing equivalent modeling methods, except for [28], where a black-box model is tested on a real microgrid. In this context, the contributions of this paper are summarized as follows.

- The modeling and identification method introduced in [34] for ADNs is adapted to microgrids.
- The so obtained method is validated with experimental data. Measurements are collected from a real LV microgrid that include a synchronous generator, two batteries, a photovoltaic (PV) power plant and static loads.
- The results of experimental tests prove the consistency and the accuracy of the proposed method, highlighting the adaptability to different microgrid configurations, including both islanded and grid-connected operating conditions.

A first version of this paper has been proposed in [39]. This new extended version includes new results, an enhanced methodology description, and further details concerning the experimental setup.

The paper is organized as follows. Section II recalls the modeling and identification methodology. Section III describes the experimental setup. Section IV presents results. Section V summarizes the conclusions of the paper.

## II. Methodology

Model definition and parameters identification represent the two main tasks of the proposed methodology, whose flowchart is illustrated in Fig. 1. The model definition is obtained through the selection of a proper model frame, in accordance with the basic a priori knowledge, required by the gray-box approach, such as the type of available sources, their rated power and operational constraints.

The parameters identification procedure defines the values of the parameters of the EDM previously selected, starting from measurements at the microgrid point of common coupling (PCC). At the end of these procedures, the EDM is fully defined, and can be used to reproduce the microgrid dynamics.

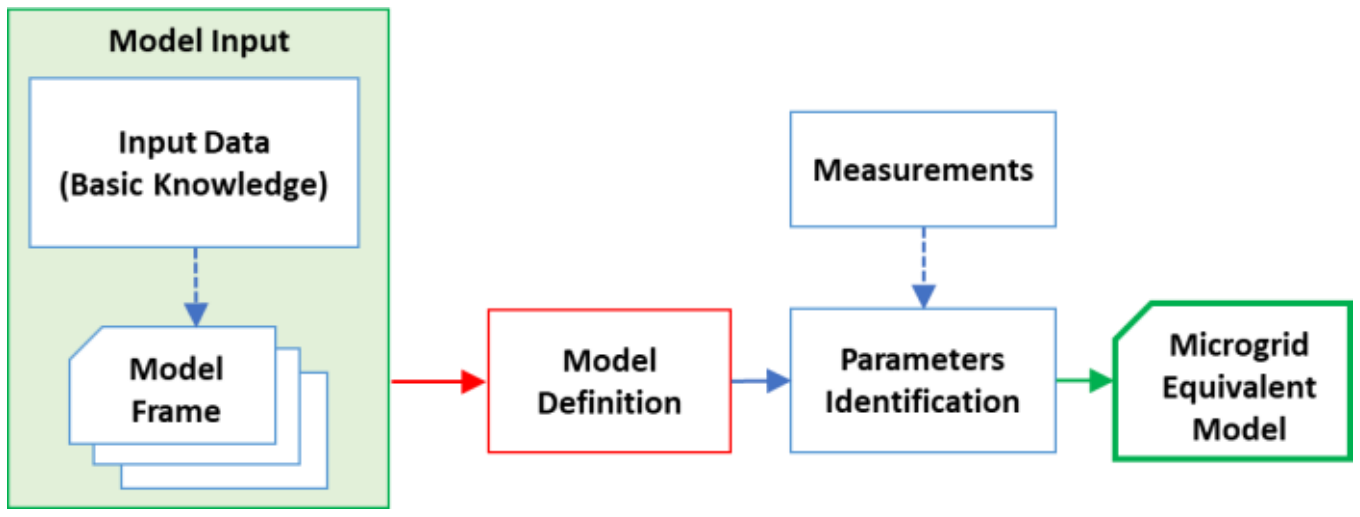

Fig. 1. Flowchart of the methodology.

### A. Model Definition

The general frame of the microgrid EDM includes a ZIP load, a synchronous machine, and a static source. The latter covers inverter driven generation, and inverter-interfaced loads. The basic system knowledge allows a first characterization of the EDM, through the proper customization of the general model frame, illustrated in Fig. 2.

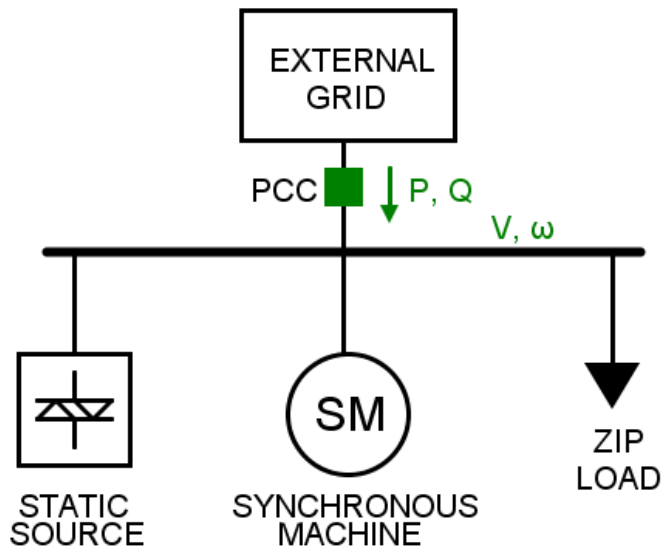

Fig. 2. General frame of the microgrid EDM.

The model can potentially include an asynchronous machine [34], which is not installed in the experimental test site considered in this paper; therefore, it has not been included in this specific application.

As shown in Fig. 2. the microgrid exchanges active and reactive powers $P$ [W] and $Q$ [var] at the PCC, where a meter is installed (green box). Positive values mean import, negative values mean export. Moreover, $v$ [pu] is the bus voltage and $\omega$ [pu] is the grid angular velocity (frequency) at the PCC.

The objective of the EDM is to reproduce $P$ and $Q$ responses to voltage and frequency variations at the PCC. Each component has been modeled as described in following.

*1) Equivalent synchronous machine (SM)*

The SM is represented with a 3rd-order dynamic model [30]:

$$\dot{e_s'} = \frac{1}{T_{ds}'}\left(e_f - \frac{x_s}{x_s'}e_s' + \frac{x_s - x_s'}{x_s'}v\cos(\delta_s)\right) \quad (1)$$

$$\dot{\omega}_s = \frac{1}{H_s}\left(t_{ms} - \frac{v_s e_s'}{x_s'}\sin(\delta_s) - D(\omega_s - \omega)\right) \quad (2)$$

$$\dot{\delta}_s = \Omega^n(\omega_s - \omega) \quad (3)$$

where: $e_s'$ [pu] is the voltage behind the transient reactance $x_s$, $\omega_s$ [pu] is the machine angular velocity, $\delta_s$ is the angle between $e_s'$ and $v$.

In order to maintain a reduced order version of the model, the governor and the exciter dynamics have not been included in the representation [34]. Therefore, the mechanical torque and the field voltage are considered as constant quantities. The active and reactive power, $P_s$ and $Q_s$, generated by the SM are given by the following relations:

$$P_s = S_s^n \frac{v e_s'}{x_s'}\sin(\delta_s) \quad (4)$$

$$Q_s = -S_s^n\left(\frac{v e_s'}{x_s'}\cos(\delta_s) - \frac{v^2}{x_s'}\right) \quad (5)$$

*2) Equivalent static source*

Active and reactive power exchanges of the static source depend on voltage magnitude and frequency deviations. These dependencies are modeled by the following linear relations:

$$P_{static} = R_P v + D_P(\omega - 1) \quad (6)$$

$$Q_{static} = R_Q v + D_Q(\omega - 1) \quad (7)$$

where parameters $R_P$, $D_P$ and $R_Q$, $D_Q$, are defined to catch the most significant behavior of static sources, equipped with droop based controllers [40]-[42].

The model has been selected in accordance with the basic system knowledge, as will be detailed in Section III-A, that highlights an instantaneous droop regulation action provided by static resources.

## B. Equivalent static load

The equivalent static load model implements the dependency of load to voltage magnitude variation as a composition of three contributions, which are the constant impedance (Z), constant current (I) and constant power (P) components (ZIP) [18]:

$$P_{load} = P_Z v^2 + P_I v + P_P \quad (8)$$

$$Q_{load} = Q_Z v^2 + Q_I v + Q_P \quad (9)$$

The model represents an equivalent static load, which reflects the real test case application.

*1) Microgrid power exchange*

The active and reactive power exchange at the PCC is obtained by the sum of the three contributions in (4)-(5), (6)-(7), and (8)-(9):

$$P = P_Z v^2 + P_I' v + P_P + D_P(\omega - 1) - S_s^n \frac{v e_s'}{x_s'}\sin(\delta_s) \quad (10)$$

$$Q = Q_Z v^2 + Q_I' v + Q_P + D_Q(\omega - 1) + S_s^n\left(\frac{v e_s'}{x_s'}\cos(\delta_s) - \frac{v^2}{x_s'}\right) \quad (11)$$

where $P_I' = P_I + R_P$ and $Q_I' = Q_I + R_Q$. Indeed, in (6)-(9) it is possible to note that both the impedance constant component of the ZIP load and the static source introduce a proportional response to the voltage variation.

It is worth remarking that the choice of the SM third-order model (1)-(3) and of the ZIP load model (8)-(9) is an option selected among others. Higher-order models of SMs are generally more accurate but their use would increase the number of parameters to be identified, and the effect on equivalent modelling could be an issue instead of an advantage. Similarly, a simple load model has been privileged. However, despite the ZIP load model appears to be quite simple, it is widely used for representing static load, which is the case of this current application. Moreover, the proposed approach has been designed to also include the presence of asynchronous machine motors [34]. The reason why the motor equivalent model has not been included in the present work is that it is not present in the current Test Facility configuration. Results provided in [34] and in previous works (*e.g.*[26], [30]-[32]) adopting these two models, show that this choice constitutes a good trade-off between complexity and accuracy. Finally, it is worth remarking that the proposed model can be further enriched to include other unmodelled components that have impact on the microgrid dynamics, such as advanced distributed controllers [9].

## C. Parameters Identification

Vector $\theta$ in (12) includes the entire set of parameters that has to be estimated by the identification procedure. It is composed by three subsets: vector $\theta_{sm}$, in (13), collects the equivalent SM parameters; vectors $\theta_v$, in (14), and $\theta_\omega$, in (15), include parameters that define the system response to voltage and frequency variations, respectively, excluding the SM dynamics:

$$\theta = [\theta_{sm} \quad \theta_v \quad \theta_\omega] \quad (12)$$

$$\theta_{sm} = [T_{ds}' \quad x_s \quad x_s' \quad H_s \quad t_{ms} \quad S_s^n \quad e_f \quad D] \quad (13)$$

$$\theta_v = [P_Z \quad P_I' \quad P_P \quad Q_Z \quad Q_I' \quad Q_P] \quad (14)$$

$$\theta_\omega = [D_P \quad D_Q] \quad (15)$$

The measurements set used in the identification process is composes by $v$, $\omega$, $P$, and $Q$ at the PCC. Assuming to have a dataset composed by $N$ samples of each quantity, the parameters estimation is obtained by solving the following nonlinear constrained optimization problem:

$$\theta^* = \min_{\theta} = \sum_{k=0}^{N-1}\left[\frac{\left(P_k - \hat{P}_{k,\theta}\right)^2}{P_0^2} + \frac{\left(Q_k - \hat{Q}_{k,\theta}\right)^2}{Q_0^2}\right] \qquad (16)$$

subject to:

$$0 < T'_{ds} \le T'_{ds,\max} \qquad (17)$$

$$0 < x'_s < x_s \qquad (18)$$

$$0 < H_s \le H_{s,\max} \qquad (19)$$

$$0 \le S_s^n \qquad (20)$$

$$0 < D \qquad (21)$$

$$0 \le t_{ms} \le 1 \qquad (22)$$

$$0 < x_s(\cos(\delta^0_{s,\max}) - e'^0_{s,\min}) + x'_s e_f - x'_s \cos(\delta^0_{s,\max}) \qquad (23)$$

$$0 \le x_s(e'^0_{s,max} - 1) - x'_s e_f + x'_s \qquad (24)$$

$$0 \le -x'_s t_{ms} + e'^0_{s,\max}\sin(\delta^0_{s,\max}) \qquad (25)$$

In (16), $P_k$ and $Q_k$ are the active and reactive power $k$-sample of measurements, and $\hat{P}_{k,\theta}$ and $\hat{P}_{k,\theta}$ are the active and reactive power obtained by simulating the microgrid EDM, given a value for the parameter set $\theta$.

Constraints (17)-(22) are modeling assumptions. All parameters are indeed forced to be positive, since they are defined as positive quantities. Moreover, $T'_{ds}$ and $H'_s$ are limited by maximum values. Constraints (23)-(25) are obtained by combining operating constraints with the steady-state equations of system (1)-(3) (apex '0' indicates steady state values). The operational constraints are that the SM field voltage $e_f$ is large enough to obtain that $e_s'^0 > 0$, (SM working as generator), and that $0 < \delta_s^0 \le \delta^0_{s,\max}$, with $\delta^0_{s,\max} \in (0,90)$ deg.

Notice that constraints (17)-(22) are all linear with respect to the model parameters in $\theta_{ms}$. Whereas (23)-(25) are quadratic but they can be easily transformed into linear inequalities by replacing $x_s$ and $x'_s$ with the auxiliary variables $\alpha_s = x_s/x'_s$ and $\alpha'_s = 1/x'_s$.

To initialize the optimization, an initial value of the model parameters should be indicated. This step is fundamental since the problem is highly nonlinear and non-convex. Thus, many local minima could exist, and a correct initialization is crucial to obtain an accurate estimation of the parameters. Initialization is carried out using all a priori information about the microgrid, such as the total nominal power of generators and loads.

Once initialized, each parameter $\vartheta_i$ is associated to a percentage confidence interval $tol^{\%}_{\vartheta_i}$, and the following constraint is added to the optimization problem:

$$\vartheta_i^0\left(1 - \frac{tol^{\%}_{\vartheta_i}}{100}\right) \le \vartheta_i \le \vartheta_i^0\left(1 + \frac{tol^{\%}_{\vartheta_i}}{100}\right) \qquad (26)$$

The parameters identification is finally realized by solving problem (16)-(26), using a proper solver for nonlinear optimization. In this work, the "interior-point" method, implemented by the MATLAB® function *fmincon*, is used.

## III. Experimental Setup

This section is dedicated to the experimental campaign details. The first part includes also the description of the test facility used to validate the equivalent modeling approach introduced in Section II.

### A. Test facility description

The Test Facility (TF) owned by RSE is a LV microgrid designed to perform studies and experimental tests on distributed energy resources (DERs) and smart grids methodologies [43]. Fig. 3 provides a schematic description of the TF.

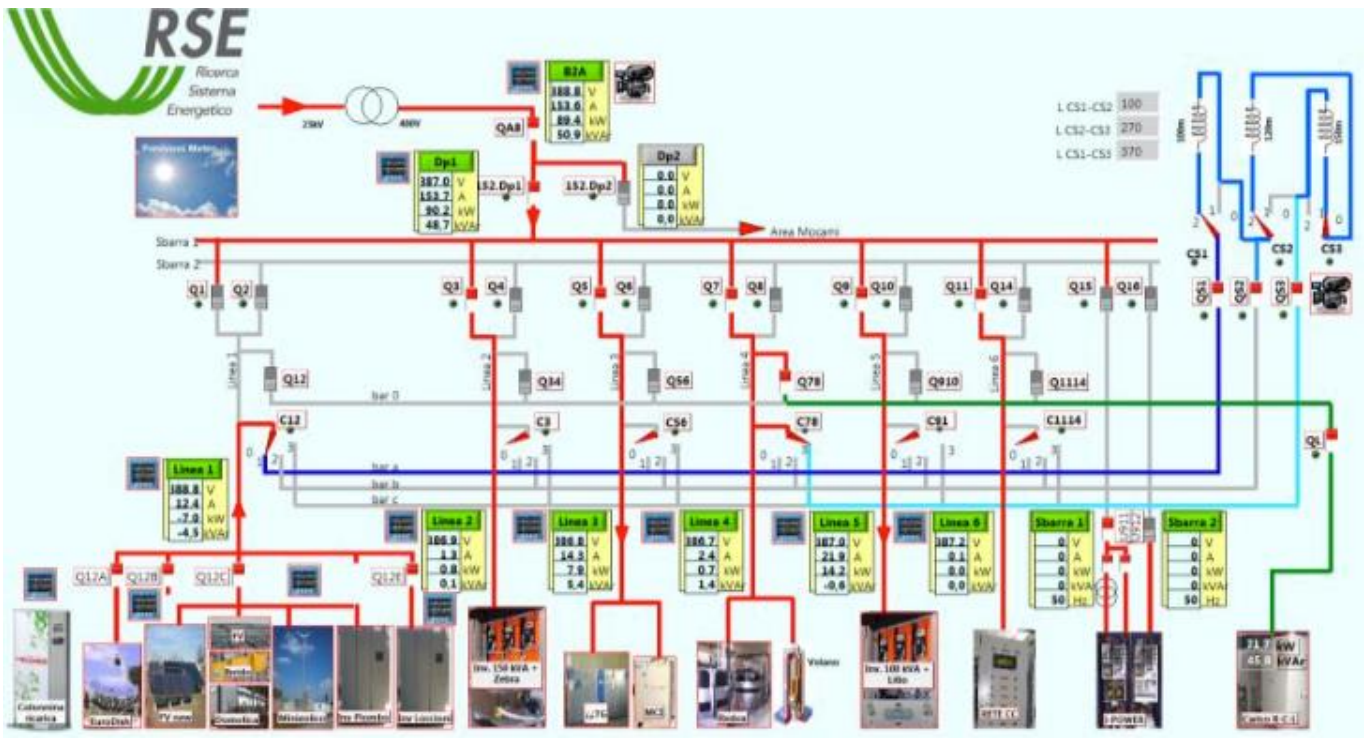

Fig. 3. Test facility description.

The network is composed of a MV/LV transformer (800 kVA) and 6 LV feeders, that can be extended using line segments (100, 150, and 200 m length).

The following generators are installed on the TF: different types of PV plant (total of 25kWp), a micro-wind generator (2kWp), and a CHP with a gas synchronous generator (50kW). Moreover, there are battery energy storage systems (BESSs) (Litium, SoNick, Vanadium Redox, Lead Acid), programmable loads (resistive, inductive and capacitive), and a 400V-100kW AC/DC interface with a DC grid.

Fig. 4 shows the block diagram and the composition of the portion of the TF used in this study, and the three configurations adopted to generate the experimental datasets, which are detailed in the next subsection III.B.

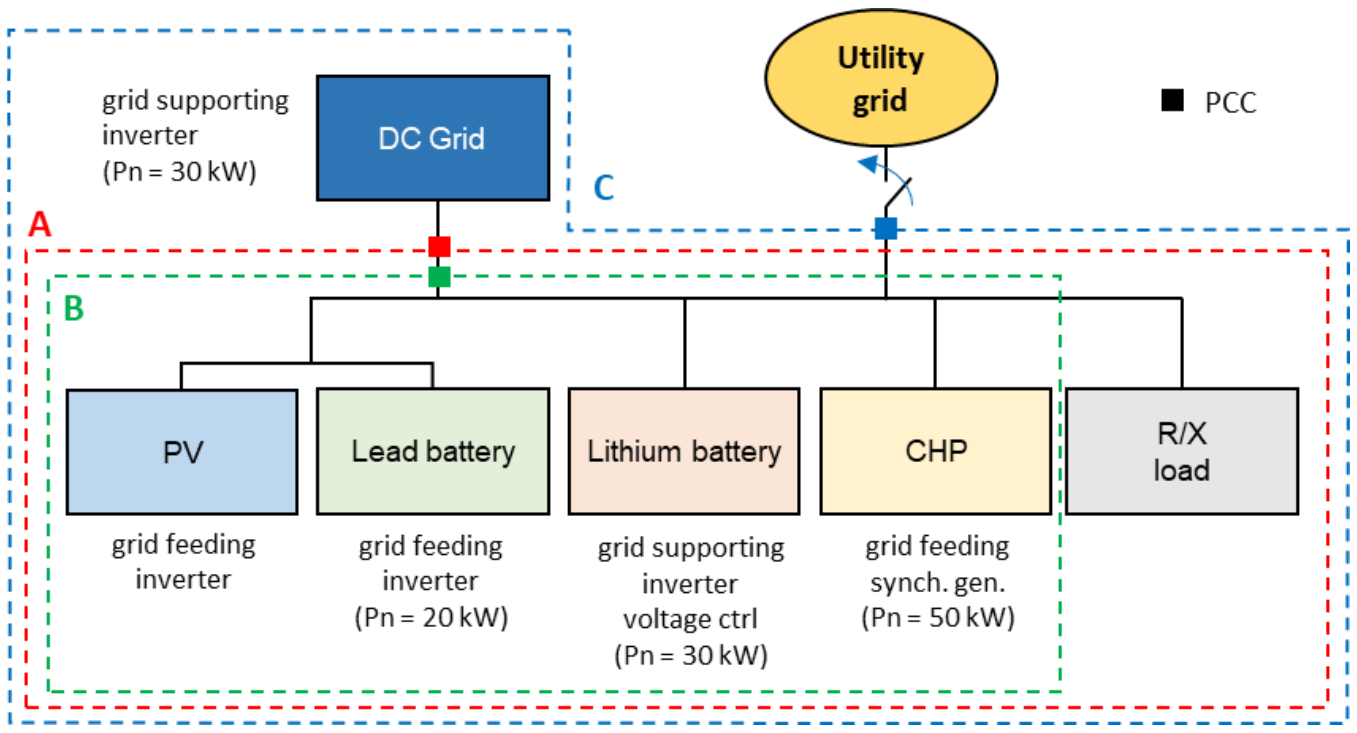

Fig. 4. Diagram of the portion of Test Facility used in this study. Dashed lines define the aggregate which is represented by the EDM for each configuration.

## B. *Experimental scenarios*

Three classes of configurations have been considered:

1) *Configuration A: TF in islanded mode with perturbations determined by the AC/DC converter*. The portion of TF considered as the system to be identified is the one within the red box in Fig. 4, the PCC is the point of connection with the DC grid, who plays the role of external grid. The AC/DC inverter operates in "grid forming" mode [44] and imposes frequency and voltage.
2) *Configuration B: TF in islanded mode with load variations*. The portion of TF considered as the system to be identified is the one within the green box in Fig. 4, the PCC is the point of connection with the DC grid.
3) *Configuration C: TF connected to the main grid*. The portion of TF considered as the system to be identified is the one within the blue box in Fig. 4, the PCC is the point of connection with the main grid.

Table I reports the details of all experimental scenarios. For each scenario, a dataset of measurements has been collected using a PMU with sampling time $T_s$ = 0.02s. Fig. 5 to Fig. 7 show three examples of datasets, one for each configuration. The identification algorithm described in Section II has been applied to the datasets collected in Scenarios 1-8. In all the cases, only the first half (in terms of time) of measurements has been used for the EDM parameters identification. Two possible initial values of parameters have been used to initialize the optimization:

- *CHP out of service (Scenarios 1-2 and 5-8)*: the initial conditions for the nominal power $S_s^{\mathrm{nom}}$ of the SM is equal to zero with a null confidence interval ($tol_{\vartheta_i}^{\%}$= 0). This means that the generator dynamics has no weight in the power exchange, and parameters to be identified are those in $\theta_v$ and $\theta_\omega$.
- *CHP in service (Scenarios 3-4)*: $S_s^{\mathrm{nom}}$ is initialized with 50kVA, which is the nominal power of the CHP. In this case, the confidence interval is zero since the information is exactly known. The mechanical torque $T_{ms}$ is 0.6p.u. and null confidence interval, since in all the cases, the CHP generates 30kW. Finally, based on an estimate provided by RSE, the inertia time constant of the CHP is within the interval [0.2 0.5] s. Thus, $H_s$ is initialized with 0.35s with 40% of confidence.

To study the capability of the identification procedure to provide an estimate of the inertia time constant, a test on Scenario 4 has been carried out using an erroneous initialization of $H_s$: 5s with a 100% confidence interval.

TABLE I. EXPERIMENTAL SCENARIOS

| | | |
|---|---|---|
| | Configuration A | |
| Scenario 1 | Li* inverter | Voltage controlled |
| | Ld* inverter | Current controlled |
| | Virtual impedance (Li) | $r$ = 3, $x$ = 3, $k_f$ = 1, $k_v$ = 5 |
| | CHP | Out of service |
| | Experiment duration | 785s |
| Scenario 2 | Li inverter | Voltage controlled |
| | Ld inverter | Current controlled |
| | Virtual impedance (Li) | $r$ = 3, $x$ = 0, $k_f$ = 1, $k_v$ = 5 |
| | CHP | Out of service |
| | Experiment duration | 752s |
| Scenario 3 | Li inverter | Voltage controlled |
| | Ld inverter | Current controlled |
| | Virtual impedance (Li) | $r$ = 2, $x$ = 5, $k_f$ = 1, $k_v$ = 5 |
| | CHP | 30 kW |
| | Experiment duration | 1623s |
| | Configuration B | |
| Scenario 4 | Li-Ld-DC inverters | Voltage controlled |
| | Virtual impedance (Li-Ld) | $r$ = 3, $x$ = 3, $k_f$ = 1, $k_v$ = 5 |
| | Virtual impedance (DC) | $r$ = 3, $x$ = 3, $k_f$ = 0.1, $k_v$ = 0.5 |
| | CHP | 30 kW |
| | Load | 27 kW |
| | Experiment duration | 394s |
| Scenario 5 | Li-Ld-DC inverters | Voltage controlled |
| | Virtual impedance (Li-Ld) | $r$ = 3, $x$ = 3, $k_f$ = 1, $k_v$ = 5 |
| | Virtual impedance (DC) | $r$ = 3, $x$ = 3, $k_f$ = 0.1, $k_v$ = 0.5 |
| | CHP | Out of service |
| | Load | 5 kW |
| | Experiment duration | 291s |
| Scenario 6 | Li-DC inverters | Voltage controlled |
| | Ld inverter | Current controlled |
| | Virtual impedance (Li-DC) | $r$ = 3, $x$ = 0, $k_f$ = 0.1, $k_v$ = 0.5 |
| | CHP | Out of service |
| | Experiment duration | 935s |
| Scenario 7 | Li-DC inverters | Voltage controlled |
| | Ld inverter | Current controlled |
| | Virtual impedance (Li-DC) | $r$ = 3, $x$ = 3, $k_f$ = 0.1, $k_v$ = 0.5 |
| | CHP | Out of service |
| | Experiment duration | 935s |
| Scenario 8 | Li-DC inverters | Voltage controlled |
| | Ld inverter | Current controlled |
| | Virtual impedance (Li-DC) | $r$ = 2, $x$ = 5, $k_f$ = 0.1, $k_v$ = 0.5 |
| | CHP | Out of service |
| | Experiment duration | 935s |
| | Configuration C | |
| Scenario 9 | Li-Ld-DC inverters | Voltage controlled |
| | Virtual impedance (Li-Ld-DC) | $r$ = 3, $x$ = 0, $k_f$ = 1, $k_v$ = 5 |
| | CHP | 30kW |
| | Load | 45kW |
| | Experiment duration | 1623s |
| Scenario 10 | Li-Ld-DC inverters | Voltage controlled |
| | Virtual impedance (Li-Ld-DC) | $r$ = 3, $x$ = 3, $k_f$ = 1, $k_v$ = 5 |
| | CHP | 30kW |
| | Load | 45kW |
| | Experiment duration | 1326s |

*Li: Lithium battery, Ld: lead acid battery, DC: AC/DC inverter

## A. *Parameters identification results*

Table II and Table III report the values of the parameters identified for Scenarios 1-8. Table II is about the parameters of the static component of the model, which have been identified for all scenarios. Table III is about the parameters of the dynamic component of the model (i.e. SM model), which have been identified only for Scenarios 3 and 4, where the CHP is in service. As above mentioned, for Scenario 4, an identification test has been carried out using an erroneous initialization of the inertia time constant $H_s$. In the two tables, and hereafter in the paper, results obtained with this erroneous initialization are indicated with “4e”.

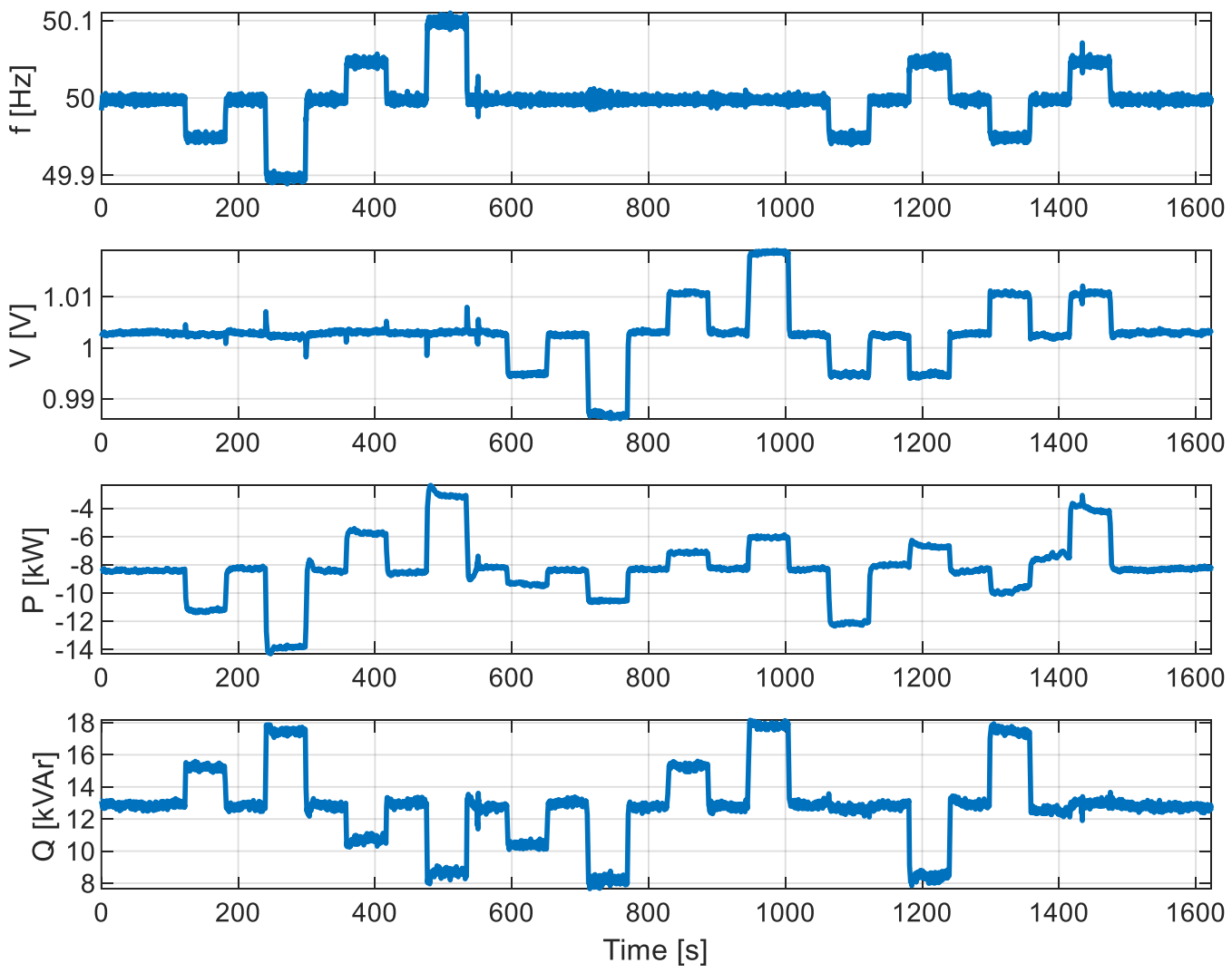

Fig. 5. Scenario 3 (Configuration A): measurements dataset

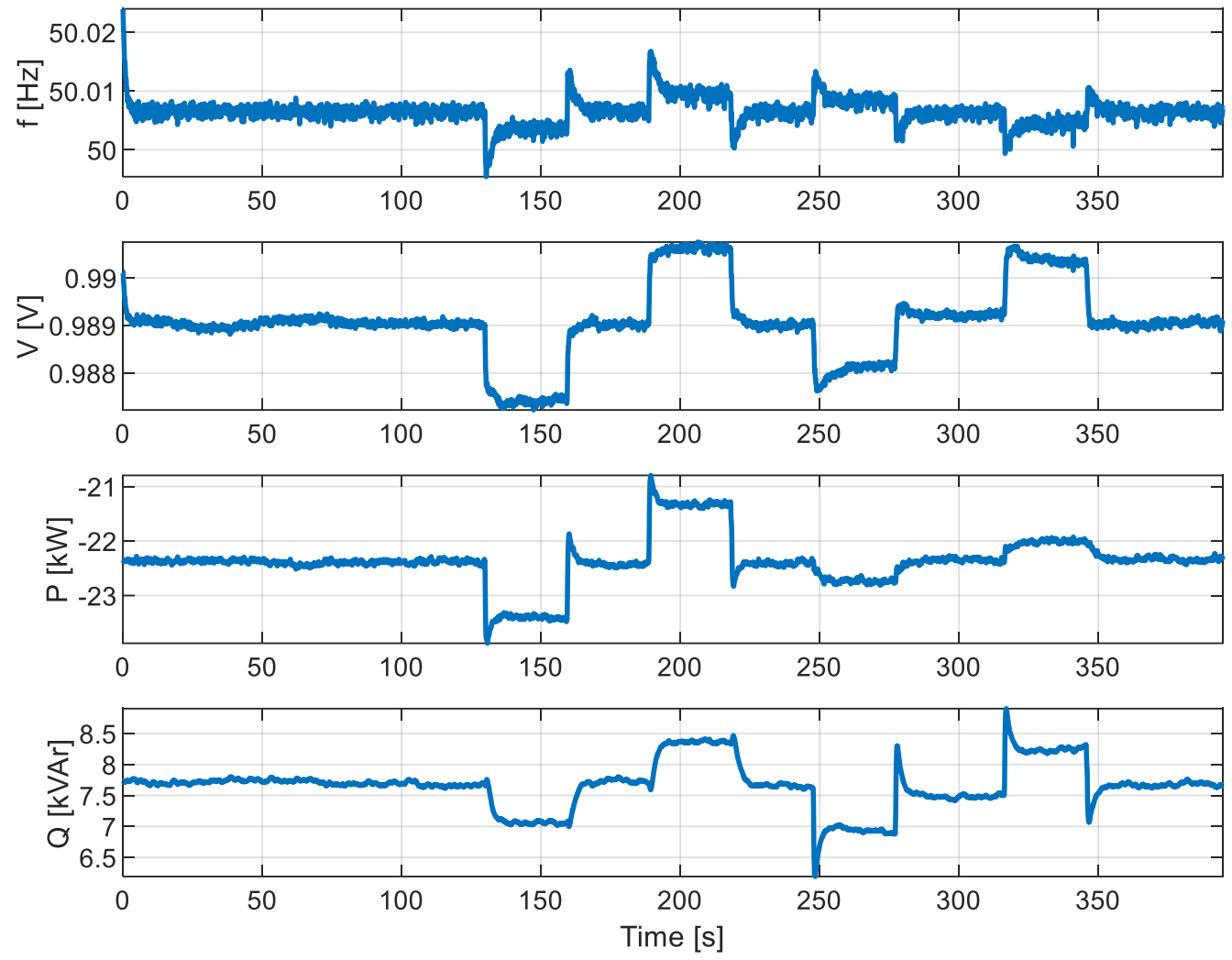

Fig. 6. Scenario 4 (Configuration B): measurements dataset.

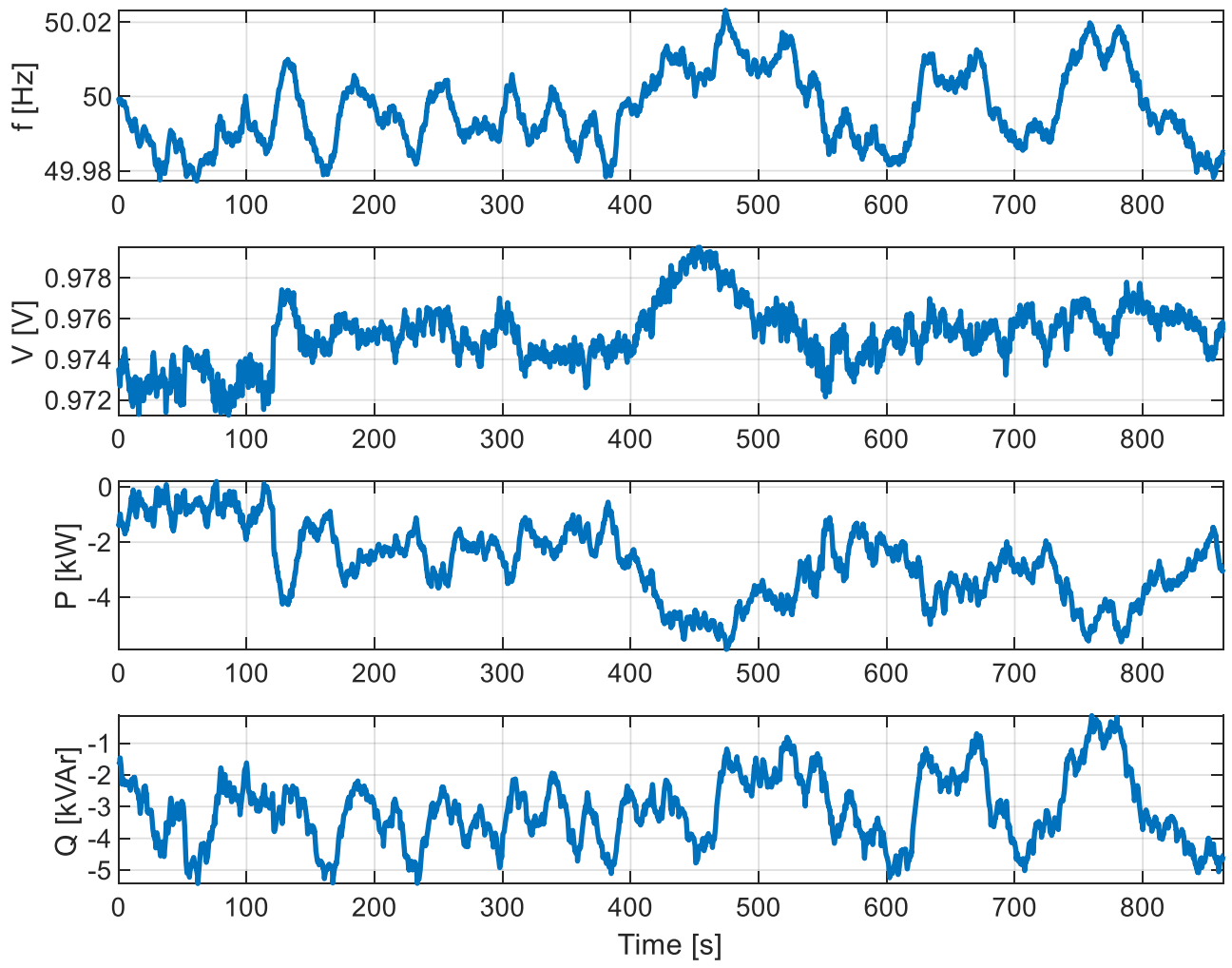

Fig. 7. Scenario 10 (Configuration C): measurements dataset.

TABLE II. IDENTIFIED VALUES OF THE PARAMETERS IN $\theta_v$ AND $\theta_\omega$

| Sc. | $P_Z$ | $P_I'$ | $P_P$ | $D_P$ | $Q_Z$ | $Q_I'$ | $Q_P$ | $D_Q$ |
|---|---|---|---|---|---|---|---|---|
| 1 | 3.31 | -6.39 | 3.07 | 1.98 | 0.00 | 0.20 | -0.20 | -2.50 |
| 2 | 0.00 | 0.30 | -0.30 | 0.53 | 0.00 | 0.01 | -0.01 | -3.14 |
| 3 | 0.30 | -0.46 | 0.18 | 2.75 | 0.59 | -0.88 | 0.32 | -2.27 |
| 4 | 0.09 | 0.35 | -0.43 | 2.43 | 0.96 | -1.44 | 0.51 | -2.50 |
| 4e | 0.09 | 0.32 | -0.43 | 2.43 | 0.98 | -1.43 | 0.52 | -2.49 |
| 5 | 0.01 | 0.49 | -0.49 | 1.98 | 0.01 | 0.42 | -0.42 | -3.14 |
| 6 | 0.00 | 0.37 | -0.38 | 0.61 | 0.30 | -0.58 | 0.28 | -2.29 |
| 7 | 1.34 | -2.40 | 1.06 | 1.44 | 0.00 | 0.27 | -0.26 | -1.95 |
| 8 | 0.07 | 0.01 | -0.09 | 1.84 | 0.17 | 0.03 | -0.20 | -1.37 |

TABLE III. IDENTIFIED VALUES OF PARAMETERS IN $\theta_{sm}$

| Sc. | $D$ | $T'_{ds}$ | $H_s$ | $S_s^{\text{nom}}$ | $t_{ms}$ | $e_f$ | $x_s$ | $x'_s$ |
|---|---|---|---|---|---|---|---|---|
| 3 | 5.01 | 0.81 | 0.25 | 0.05 | 0.6 | 2.13 | 2.60 | 0.19 |
| 4 | 5.70 | 0.76 | 0.26 | 0.05 | 0.6 | 2.12 | 2.20 | 0.18 |
| 4e | 4.75 | 0.72 | 0.20 | 0.05 | 0.6 | 2.15 | 2.49 | 0.11 |

In Table II and Table III, it is possible to note that the values of parameters of the SM identified within the different scenarios are very similar each other. This means that the identification algorithm can uniquely detect the dynamical behavior of the CHP. However, it is worth remarking that the values computed with the erroneous initialization are different with respect to the ones calculated with the correct initialization, even if they have the same order of magnitude. On the one hand, this is not surprising since the model is nonlinear, and a full identifiability cannot be obtained. On the other hand, this does not mean that the model is not able to accurately reproduce the network dynamics but only that, in some scenarios, the estimation of parameters cannot be precisely carried out.

## B. *Identification and validation results*

Fig. 8 and Fig. 9 show the results of identification and validation obtained in Scenario 3 and Scenario 4, respectively. The measured profiles of $P$ and $Q$ are compared with the ones reproduced by the EDM. It is worth recalling that the first half (in terms of time) of the measurements is used in the identification procedure, whereas the rest is used for validation. Observing the figures, in both the scenarios, the EDM appears to be able to represent microgrid dynamics with a good accuracy.

Fig. 10 reports the identification and validation results obtained with the erroneous initialization in Scenario 4. As it appears clear, the EDM 4e shows an accuracy of representation very close to the one of EDM 4 (Fig. 8).

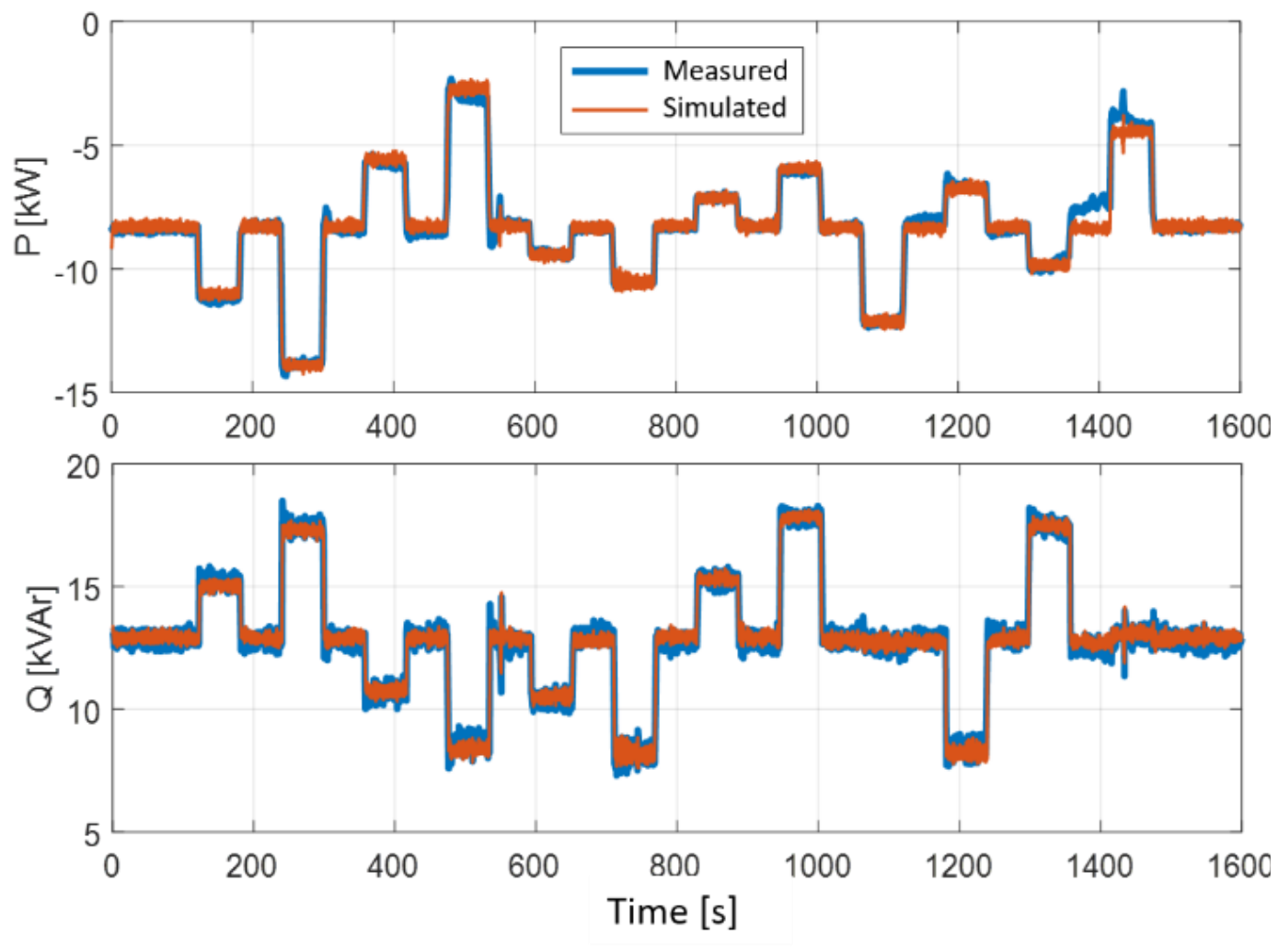

Fig. 8. Scenario 3 (Configuration A): indentification and validation results.

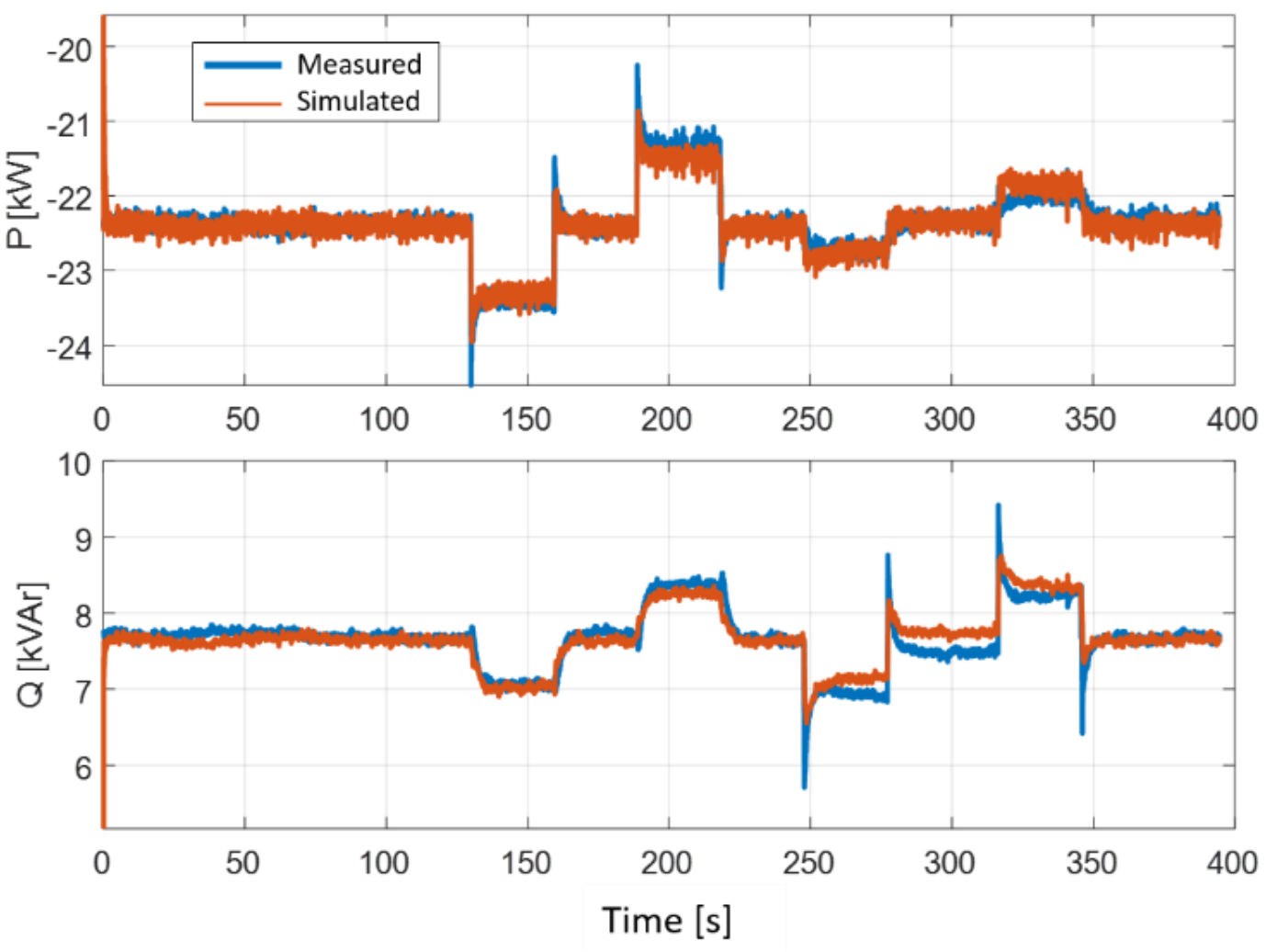

Fig. 9. Scenario 4 (Configuration B): identification and validation results.

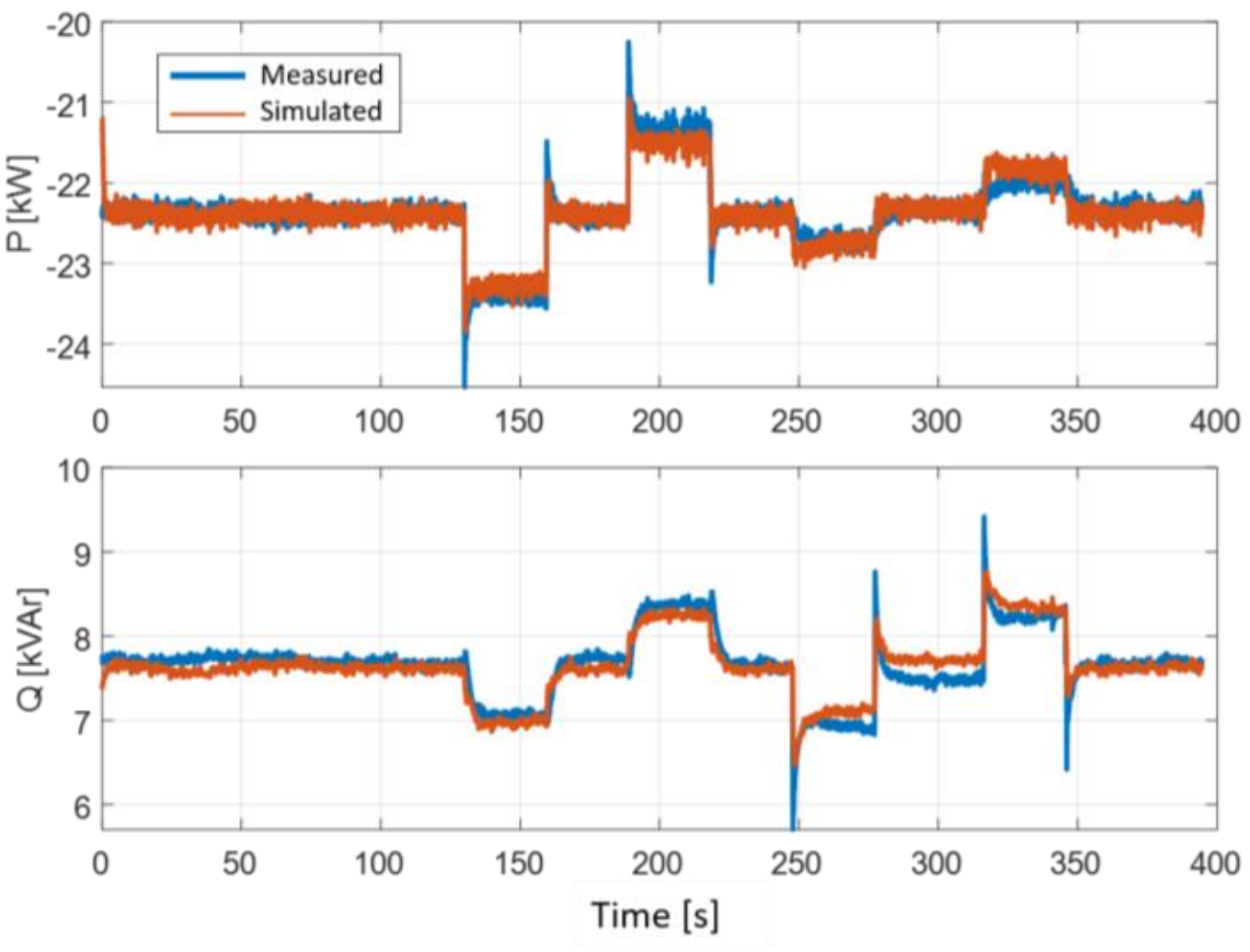

Fig. 10 Scenario 4 (Configuration B): validation results with erroneous initialization (EDM 4e).

Power profiles in Fig. 8, Fig. 9, and Fig. 10 are provided as examples of identification and validation results. To quantify the accuracy of the EDM in representing the system dynamics, the following root mean square errors (RMSEs) are computed:

$$\sigma_P = \sqrt{\frac{1}{N}\sum_{k=0}^{N-1}\left(P_k - \hat{P}_k\right)^2} \tag{23}$$

$$\sigma_Q = \sqrt{\frac{1}{N}\sum_{k=0}^{N-1}\left(Q_k - \hat{Q}_k\right)^2} \tag{24}$$

where $P_k$ and $Q_k$ are active and reactive powers measured at the sampling time $k$, and $\hat{P}_k$ and $\hat{Q}_k$ are the active and reactive powers reproduced by the EDM, respectively.

Table IV reports the RMSEs obtained in Scenarios 1-8. The item "Identification", indicates the RMSEs obtained with the first time half of datasets; the item "Validation", reports the RMSEs obtained with the second time half of datasets, not used in the identification.

TABLE IV. RMSES FOR IDENTIFICATION AND VALIDATION

| | Identification | | Validation | |
|---|---|---|---|---|
| Sc. | $\sigma_P$ [kW] | $\sigma_Q$ [kvar] | $\sigma_P$ [kW] | $\sigma_Q$ [kvar] |
| 1 | 0.388 | 0.437 | 0.510 | 0.352 |
| 2 | 0.227 | 0.630 | 0.238 | 0.534 |
| 3 | 0.290 | 0.210 | 0.295 | 0.223 |
| 4 | 0.173 | 0.146 | 0.181 | 0.149 |
| 4e | 0.180 | 0.145 | 0.183 | 0.152 |
| 5 | 0.101 | 0.079 | 0.125 | 0.078 |
| 6 | 0.109 | 0.114 | 0.171 | 0.125 |
| 7 | 0.162 | 0.141 | 0.169 | 0.126 |
| 8 | 0.221 | 0.134 | 0.212 | 0.126 |

In scenarios 3-8, RMSEs are always lower than 0.3kW for $P$ and 0.2kvar for $Q$. In Scenarios 1-2, RMSEs are lower than 0.5kW for active power and 0.6kvar for reactive power. This consideration holds true both for identification and validation. In all the scenarios, the order of magnitude of active and reactive power variations are lower than 1kW and 1kvar. Therefore, errors can be considered sufficiently low, meaning that the EDM is able to reproduce the microgrid active and reactive power responses with a good accuracy.

Notice also that the difference between the RMSEs obtained in identification and validation is lower than 0.1kW and 0.1kvar in Scenarios 1-2 and of the order of centimes of kW and kvar in Scenarios 3-8. Finally observe that RMSEs obtained in Scenarios 4 and 4e are very close each other.

### *C. Cross-validation results*

The final part of the analysis is dedicated to the consistency and scalability evaluation of the proposed EDM. The two EDMs identified with the dataset of Scenario 4, with correct and incorrect initializations (models 4 and 4e) have been applied to the other scenarios where the CHP is in service, *i.e.* Scenario 3 and Scenarios 9 and 10. Notice that in these two last cases, differently from all other scenarios, the TF is connected to the main grid (Configuration C).

Fig. 11 and Fig. 12 show two examples of the obtained results. In Fig. 11, the power profiles measured in Scenario 3 are reproduced by the EDM 4. By comparing such profiles with the ones in Fig. 8, where the EDM 3 is employed, no significant differences can be appreciated. Fig. 12 reports the results obtained with model EDM 4e, applied to Scenario 10. Also in this case, the EDM appears to be capable to correctly reproduce the microgrid dynamic response.

Table V reports the cross-validation RMSEs. These values are not significantly different from the ones obtained for identification and validation (see Table IV). By cross-validation results, it is possible to conclude that the EDM shows a good accuracy of representation and the approach is consistent and scalable.

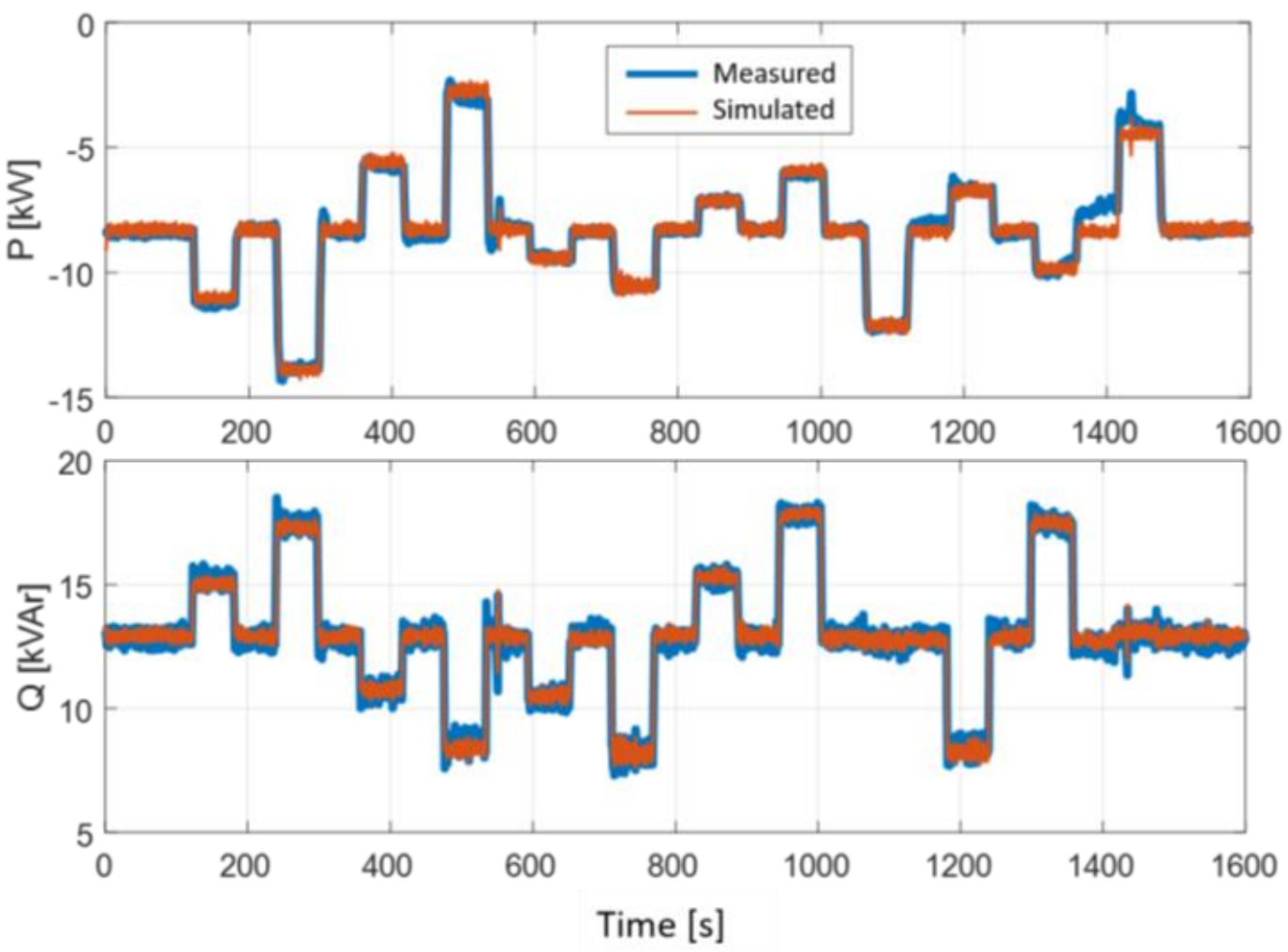

Fig. 11. Cross-validation results. Active and reactive power measured in Scenario 3 and reproduced with the EDM indentified with the dataset of Scenario 4 (EDM 4).

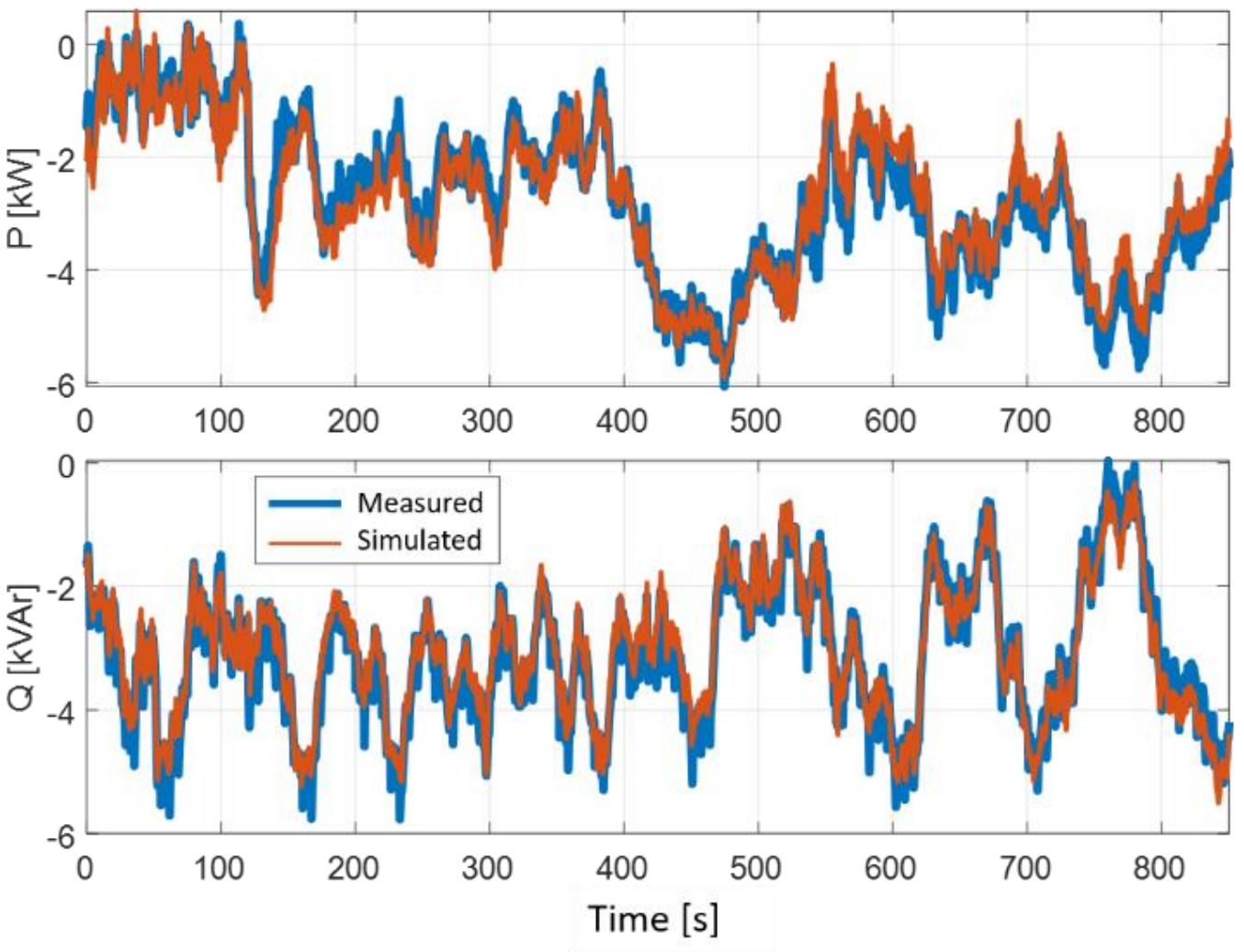

Fig. 12. Cross-validation results. Active and reactive power measured in Scenario 10 and reproduced with the EDM indentified with the dataset of Scenario 4 with erroneous initialization (EDM 4e).

TABLE V. RMSEs FOR CROSS-VALIDATION

| Scenario | Model | $\sigma_P$ [kW] | $\sigma_Q$ [kvar] |
|---|---|---|---|
| 3 | 4 | 0.290 | 0.210 |
| 3 | 4e | 0.290 | 0.211 |
| 9 | 4 | 0.488 | 0.226 |
| 9 | 4e | 0.488 | 0.233 |
| 10 | 4 | 0.359 | 0.236 |
| 10 | 4e | 0.356 | 0.235 |

Moreover, cross-validation results are useful to show the potential capability of the EDM to be used in real applications. Indeed, models EDM 4 and EDM 4a have been identified, for a configuration with the CHP in service, through a procedure internal to the microgrid. Then they have been used with success to model the microgrid response in the grid-connected mode. This can be applied for all possible configurations of the microgrid to build a set of EDMs, and to select the right one in real-time as a function of the current configuration. The latter can be known if triggered by advanced controllers which activate or deactivate the different microgrid devices or detected by a suitable monitoring system.

## IV. CONCLUSIONS

In this paper, the equivalent modeling technique developed in [34] has been adapted to microgrids and validated by experimental tests. Specifically, measurements of voltage, frequency, and active and reactive powers have been collected from a real LV microgrid, that include a synchronous generator, two batteries, a photovoltaic (PV) power plant and static loads. Three different configurations of the system, for a total of ten scenarios, have been considered.

The obtained results show that the proposed EDM is able to accurately reproduce the dynamic response of the microgrid to external disturbances. In particular, given a variation on voltage and frequency at the PCC, the EDM correctly returns the variations of active and reactive power exchanged by the microgrid. The method has resulted to be robust with respect to a less accurate initialization. Moreover, cross-validation tests have shown that the EDM can be adapted without difficulties to different configurations.

Future works will consider the extension of the method with an automatic update of the model after microgrid configuration changes.

## ACKNOWLEDGMENTS

This work has been financed by the Research Fund for the Italian Electrical System, Italy in compliance with the Decree of April 16, 2018.